\documentclass[%
 twocolumn,
superscriptaddress,
 amsmath,amssymb,
 prb,
]{revtex4-1}

\usepackage{graphicx}% Include figure files
\usepackage{dcolumn}% Align table columns on decimal point
\usepackage{bm}% bold math
\usepackage[all]{xy}
\usepackage{color}
\usepackage{soul}
\begin{document}

%\preprint{APS/123-QED}

\title{String monopoles, string walls, vortex skyrmions and nexus objects in polar distorted B-phase of $^3$He}% Force line breaks with \\

\author{G.E.~Volovik}
\email{grigori.volovik@aalto.fi}
\affiliation{Low Temperature Laboratory, Aalto University,  P.O. Box 15100, FI-00076 Aalto, Finland}
\affiliation{Landau Institute for Theoretical Physics, acad. Semyonov av., 1a, 142432,
Chernogolovka, Russia}

\author{K.~Zhang}%
\email{kuang.zhang@aalto.fi}
\affiliation{Low Temperature Laboratory, Aalto University,  P.O. Box 15100, FI-00076 Aalto, Finland}
\affiliation{University of Helsinki, Department of Mathematics and Statistics, P.O. Box 68 FI-00014, Helsinki, Finland}

\date{\today}% It is always \today, today,

\begin{abstract}
The composite cosmological objects --  Kibble-Lazarides-Shafi (KLS) walls bounded by strings and cosmic strings terminated by Nambu monopoles -- could be produced during the phase transitions in the early universe.
Recent experiments in superfluid $^3$He reproduced the formation of the KLS domain walls, which opened the new  arena  for the detailed study of those objects in human controlled system with different characteristic lengths. These composite defects are formed by two successive symmetry breaking phase transitions. In the first transition 
the strings are formed, then in the second transition the string becomes the termination line of the KLS wall. In the same manner, in the first transition monopoles are formed, and then in the second transition these monopoles become the termination points of strings. Here we show that in the vicinity of the second transition the composite defects  can be described by relative homotopy groups.   This is because there are two well separated length scales involved, which give rise to two different classes of the degenerate vacuum states, $R_1$ and $R_2$, and the composite objects correspond to the nontrivial elements of the group $\pi_n(R_1,R_2)$. We discuss this on example of the so-called polar distorted B-phase, which is formed in the two-step phase transition in liquid $^3$He distorted by aerogel. In this system the string monopoles terminate spin vortices with even winding number, while KLS string walls terminate on half quantum vortices. In the presence of magnetic field, vortex skyrmions are formed, and the string monopole transforms to the nexus.  We also discuss the integer-valued topological invariants of those objects. Our consideration can be applied to the composite defects in other condensed matter and cosmological systems.
\end{abstract}
\pacs{
}
\maketitle

\section{Introduction} 

The combined topological objects, such as strings terminated by Nambu monopoles \cite{Nambu1977} and Kibble-Lazarides-Shafi (KLS) walls bounded by strings \cite{Kibble1982,Kibble2000}, typically appear at two successive symmetry breaking phase transitions \cite{Kibble2015}.  Example is provided by the axion solution to the strong CP problem \cite{Vilenkin1982,Sikivie1982,Andrea2019,Chatterjee2019}, where  two different phase transitions occur as the Universe cools down. In the first one the cosmic strings are formed. Then when the cosmic temperature reaches the QCD scale, domain walls are formed, while cosmic strings become attached to these walls.

Similar formation of the combined objects in successive phase transitions has been observed in the nanoconfined superfluid $^3$He (in the so-called nafen) \cite{Makinen2019}. The confined geometry provides variety of new phenomena in this spin-triplet $p$-wave superfluid. Some new phases, which can  never be stable in bulk $^3$He, are stabilized by the nano-confinement. Among them the polar phase has been predicted \cite{Kazushi2006,Ikeda2014} and experimentally identified \cite{Askhadullin2012}. 
Later it became clear that the main reason of the dominations of the polar phase in nafen  is the extension of the Anderson theorem \cite{Anderson1959} to the polar phase with columnar impurities: the  transition temperature to the polar phase is practically not suppressed by the strands of nafen \cite{Fomin2018,Fomin2020,Ikeda2019b}, as distinct from the other superfluid phases.  Similar extension of  the Anderson theorem was also discussed in connection with multi-orbital superconductors \cite{Ramires2018}.
Another signature of the Anderson theorem  is the observation that the Dirac nodal line in the spectrum of Bogoliubov quasipartilces in the polar phase is not suppressed by nafen strands \cite{Eltsov2019}, giving rise to the detected $T^3$ dependence of the gap amplitude.

In the polar phase the half quantum vortices (HQVs) have been observed \cite{Autti2016}. Originally the HQVs were predicted to exist in the chiral superfluid $^3$He-A \cite{VolovikMineev1976,Cross1977}, but for forty years they escaped observation because in this phase they are  energetically unfavorable compared to the singly quantized vortices. Now the HQVs are easily created in $^3$He-A: the HQVs are first formed in the polar phase, where they are energetically favorable, and then they survive after the phase transition from polar phase to the A-phase due to strong pinning \cite{Makinen2019}. The structure, spin dynamics and spin polarization of HQV in the A-phase were theoretically studied during the last few decades \cite{Salomaa1985,Hu1987,Vakaryuk2009}. Both the polar phase and   $^3$He-A are superfluids with the so-called equal spin pairing \cite{VollhardtWolfle1990}. In such systems the HQV can be considered as the vortex in one spin component, and in the chiral superfluid $^3$He-A such vortex contains single (isolated) Majorana zero mode in its core \cite{Volovik1999,ReadGreen2000,Ivanov2001}.

 Both in the bulk B-phase and in the polar distorted B-phase in nafen, the HQVs are not supported by topology.
However, earlier it was shown that the non-axisymmetric core of  singly quantized vortex 
\cite{Thuneberg1986,VolovikSalomaa1985}, which was observed in bulk superfluid $^3$He-B \cite{Kondo1991}, can be considered as a pair of HQVs connected by the KLS wall \cite{Volovik1990,Silaev2015}. This wall is rather short: only of few coherence length size. 
The similar wall between HQVs appears in the polar distorted B-phase in nafen after transition from the polar phase. But  due to strong vortex pinning in nafen, the KLS wall bounded by pinned vortices does not shrink and keeps its macroscopic size. This allows experimental detection and identification of such composite objects \cite{Makinen2019}.

Strictly speaking,  the wall bounded by strings is not a topological object, since after the second transition the topological charge of the string does not exist anymore. In the same way  the string terminated by monopole is not topological. Here we show that under certain conditions these combined objects become topological, being described by relative homotopy groups. Originally the classification  
in terms of the relative homotopy groups has been used if there is the hierarchy of the energy scale or length scales in physical system \cite{MineyevVolovik1978,Mermin1979,Michel1980}, when each energy scale has its own well defined vacuum manifold $R_i$ -- the space of the degenerate states. 
 Here are some examples:

 (i) The spin-orbit interaction in superfluid $^3$He is small, and we have the order parameter vacuum manifold $R_1$ at short distances where the spin-orbit interaction can be neglected, and the submanifold $R_2\subset R_1$ at large distances, where the space of the order parameter is restricted by spin-orbit interaction \cite{MineyevVolovik1978}.
The relative homotopy groups $\pi_n(R_1,R_2)$ give different types of the topologically stable combined  objects.
The  $\pi_1(R_1,R_2)$ describes the planar topological solitons terminated by strings. Examples are
combined spin-mass vortices with soliton tail observed in superfluid $^3$He-B \cite{Kondo1992}, and solitons terminated by half-quantum vortices observed in spinor Bose condensate with quadratic Zeeman energy \cite{Seji2019,Liu2020}. The $\pi_2(R_1,R_2)$ describes linear topological solitons and skyrmions terminated  by monopoles.

(ii) Another example of the  two-manifold  system is when the boundary conditions restrict the order parameter on the boundary, with $R_1$ being the space in bulk and $R_2\subset R_1$ is subspace on the boundary restricted by the boundary conditions. This gives the topological classification of the topological objects on the surface of an ordered system \cite{Volovik1978}, such as boojum \cite{Mermin1981}.

(iii) The  two-scale system also emerges when there is the hidden symmetry, which may soften the cores of topological defects \cite{MMV1978}. 

In our case of two successive transitions, two energy scales arise in the vicinity of the second transition. There, the coherence length related to the first symmetry breaking $G\rightarrow H_1$ is much smaller than the coherence length related to the second symmetry breaking $H_1\rightarrow H_2$. This gives rise to two well defined vacuum manifolds, $R_{1} \cong G/H_{2}$ and $R_2 \cong H_1/H_2$, and allows us to apply the  relative homotopy groups $\pi_n(R_1,R_2)$ for classification of the combined objects:  string-monopole objects (analogs of Nambu monopoles); wall-string objects (analogs of KLS wall); nexus \cite{Cornwall1999},
  etc. That is because the order parameter fields are mapped into different degenerate vacuum manifolds at different spatial regions, thus the homotopy classes of order parameters constitute  $\pi_n(R_1,R_2)$ \cite{GoloMonastyrky1978}, see details in appendices Sec. \ref{app:derivation}. In superfluid $^3$He, these topological objects live in the vicinity of transition between the polar phase and the polar distorted B-phase (PdB). 

This paper is organized as follows. In Sec. \ref{SecConventional} we consider the conventional scheme of the symmetry breaking and the vacuum manifolds of different superfluid phases appeared in the successive transitions. The topological defects in these phases, which emerge due to symmetry breaking and are described in terms  of the conventional homotopy groups of vacuum manifolds, are  considered in Sec. \ref{SecConventional2}. In Sec. \ref{SecCombined1} we discuss combined topological objects in the vicinity of the second transition, where the order parameter could be mapped into two different vacuum manifolds with different coherence lengths. We use the relative homotopy groups and corresponding  exact sequence of homomorphisms to describe the classes of combined objects, which are topologically stable in the vicinity of the transition. Based on the exact sequence of the homotopy groups, we find the topological stability of the string monopole (string terminated on monopole) and of the KLS string wall (KLS wall bounded by string). The latter has been observed in recent experiments \cite{Makinen2019}.
In Sec. \ref{VorticesSkyrmion} we discuss vortex skyrmions emerging in the presence of  magnetic field, and the nexus object.  In Sec. VI we summarize our results and discuss the role of these objects in formation of the numerous supefluid glass states, which may exist in aerogel \cite{Volovik1996,Dmitriev2010,Glasses2019}.

\section{\label{SecConventional}Conventional symmetry breaking scheme and vacuum manifolds}

The continuous phase transition is understood as spontaneous symmetry breaking by order parameters about a primary symmetry group $G$. In $^3$He liquid at low temperature, the order parameters space  consists of two 3-dimensional vector spaces and the phase space. Thus the order parameter is represented by the complex-valued dyadic tensor $A_{{\alpha}i}$ \cite{VollhardtWolfle1990}, which transforms under the action of spin, orbital and phase rotations of the group  $G$. Stabilizer of those actions forms the residual symmetry group $H$ of superfluid phase of $^3$He. In our case, the symmetry group  $G$ of  normal liquid $^3$He in the "nematically ordered" aerogel with the uniaxial anisotropy is different from that in the bulk  $^3$He \cite{VollhardtWolfle1990}. The uniaxial anisotropy in the orientation of aerogel strands in nafen reduces the symmetry under the $SO_L(3)$ group of rotations in the orbital space  to the  $O_{L}(2)$ subgroup \cite{Makinen2019}. 
If the tiny spin-orbit interaction is neglected, the normal phase vacuum has the following symmetries:
\begin{equation}
%G \cong 
O_L(2)\times SO_S(3)\times U(1) \times T \times P
 \,,
\label{Ggeneral}
\end{equation}
where
$SO_{S}(3)$ is the group of spin rotations; $U(1)$ is the global gauge group, which is broken in superfluid states; $T$ is time reversal symmetry; $P$ is parity; $O_{L}(2) \cong SO_{L}(2)\rtimes C_{2x}^L $ where $C_{2x}^L $ is $\pi$ rotation in orbital space. 

In what follows, we ignore the time reversal symmetry, since it is not broken in the polar and in PdB phases, and also ignore the parity $P$ which is reduced to $Pe^{i\pi}$ in all $p$-wave superfluid phases, where $e^{i\pi}$ is the $\pi$-rotation in phase space. Also, because we focus on the topological objects related to the spin and $U(1)$ gauge parts of the order parameter, the $\mathbb{Z}_{2}$ symmetry coming from $C_{2x}^L$ could be  neglected in the rest parts. Then the  relevant starting group $G$ of symmetry breaking scheme in this paper is
\begin{equation}
G \cong SO_L(2)\times SO_S(3)\times U(1)
 \,.
\label{G}
\end{equation}
Starting from this normal phase vacuum, we discuss three types of phase transition: (a) from the normal phase to the polar phase; (b)  from the polar phase to the PdB phase;
and (c) the possible direct transition  from the normal phase to the PdB phase. In this Section we consider the topological objects related to these symmetry breaking scenarios, using the conventional homotopy group approach.

\subsection{Transition from normal phase to polar phase}

The order parameter in the $p$-wave spin-triplet superfluids is the dyadic tensor $A_{{\alpha}i}$, which transforms as a vector under spin rotation  (the greek index) and as a vector under orbital rotations (the latin index). In the polar phase it has the form:
\begin{equation}
A^{P}_{{\alpha}i}=\Delta_P \hat{d}_{\alpha} \hat{z}_{i}e^{i\Phi}
 \,,
\label{AP}
\end{equation}
where $\Phi$ is the phase, $\hat{d}_{\alpha}(\equiv {\hat{\mathbf{d}}})$ and $\hat{z}_{i}$ are unit vectors of spin and orbital uniaxial anisotropy  respectively, 
and $\Delta_P$ is the gap amplitude. The residual symmetry group of the polar phase, the symmetry group of the order parameter (\ref{AP}),  is
\begin{equation}
H_{\rm P} \cong SO_L(2)\times SO_S(2) \rtimes \mathbb{Z}_{2}^{S-{\Phi}} \subset G
 \,.
\label{HP}
\end{equation}
Here $\mathbb{Z}_{2}^{S-{\Phi}} \cong \{1,C_{2x}^Se^{i\pi}\}$, where $C_{2x}^S$ is $\pi$-rotation of the  vector $\hat{\mathbf{d}}$ about perpendicular axis and $e^{i\pi}$ is  the phase rotation by $\pi$, i.e.  $\Phi \rightarrow \Phi +\pi$. Then the vacuum manifold of the polar phase is given as
\begin{equation}
R_{\rm P} \cong G/H_{\rm P} \cong (S^2\times  U(1))/{\mathbb{Z}}_2 %\cong \mathbb{R} P^{2} \times  U(1)
\,.
\label{RP}
\end{equation}
The coherence length $\xi =v_F/\Delta_P$ in the polar phase is the smallest length scale in our problem, which determines the size of singular (hard core) topological defects in the polar phase.

\subsection{From polar phase to PdB phase}

Let us now consider the second symmetry breaking phase transition: from the polar phase vacuum with fixed $\hat{\mathbf{d}}$ and $\Phi$ to the PdB phase.  In the vicinity of this transition the order parameter  (\ref{AP}) acquires the symmetry breaking term with amplitude $q\ll 1$:
\begin{equation}
A^{PdB}_{{\alpha}i}=e^{i\Phi}\Delta_P[\hat{d}_{\alpha} \hat{z}_{i} + q (\hat{\mathbf{e}}^{1}_{\alpha} \hat{x}_{i} + \hat{\mathbf{e}}^{2}_{\alpha} \hat{y}_{i})]
 \,.
\label{ApdbP}
\end{equation}
Here $\hat{\mathbf{e}}^{1}$, $\hat{\mathbf{e}}^{2}$ and $\hat{\mathbf{d}}$ form the triad of orthogonal vectors in spin space. The corresponding coherence length of the second transition $\xi/q$ is large in the vicinity of this transition. This provides the hierarchy of the length scales, $\xi$ and $\xi/q\gg \xi$. 

The residual symmetry subgroup of the PdB phase in the symmetry breaking from the polar phase is
\begin{equation}
H_{\rm PdB} \cong SO_{J}(2) \subset H_{\rm P} 
 \,,
\label{HPdB}
\end{equation}
where $SO_{J}(2)$ represents the common rotations of spin and orbital spaces.
 The manifold of the vacuum states, which characterizes the second symmetry breaking is:
\begin{widetext}
\begin{equation}
R_2 \equiv R_{{\rm P}\rightarrow {\rm PdB}} \cong
H_{\rm P}/H_{\rm PdB} \cong (SO_L(2)\times SO_S(2)\rtimes \mathbb{Z}_{2}^{S-{\Phi}} )/ SO_{J}(2) \cong SO_{L-S}(2) \times  \mathbb{Z}_{2}^{S-\Phi}
 \,.
\label{PtoPDB}
\end{equation}
\end{widetext}
Here $SO_{L-S}(2)$ is the broken symmetry with respect to relative rotations of spin and orbital spaces.

\subsection{From normal phase to PdB phase}

Here we consider the situation   deep  inside the PdB phase, where the parameter $q$ is not necessarily small. In this general case there is only a single length scale which  is relevant, and thus this situation becomes similar to that of the direct transition from the normal state to the PdB phase, $G\rightarrow H_{\rm PdB}$. The order parameter Eq.   (\ref{ApdbP}) of the PdB phase could be written as
\begin{equation}
 A^{PdB}_{{\alpha}i} = e^{i\Phi}[\Delta_\| \hat{d}_{\alpha} \hat{z}_{i} + \Delta_{\bot} (\hat{\mathbf{e}}^{1}_{\alpha} \hat{x}_{i} + \hat{\mathbf{e}}^{2}_{\alpha} \hat{y}_{i})]
 \,,
\label{ApdbPa}
\end{equation}
where $\Delta_{\bot}\leq\Delta_{\|}$. The corresponding residual symmetry subgroup is in Eq. (\ref{HPdB}),
and the vacuum manifold of PdB phase in this scenario  of symmetry breaking is:
\begin{equation}
R_1 \equiv R_{{\rm normal}\rightarrow {\rm PdB}} \cong G/H_{PdB} \cong SO_{L-S}(3)\times U(1)
 \,.
\label{RPdB}
\end{equation}

%%%%%%%%%%%%%
\begin{figure*}[htbp] 
  \centerline{\includegraphics[width=11cm,height=8cm]{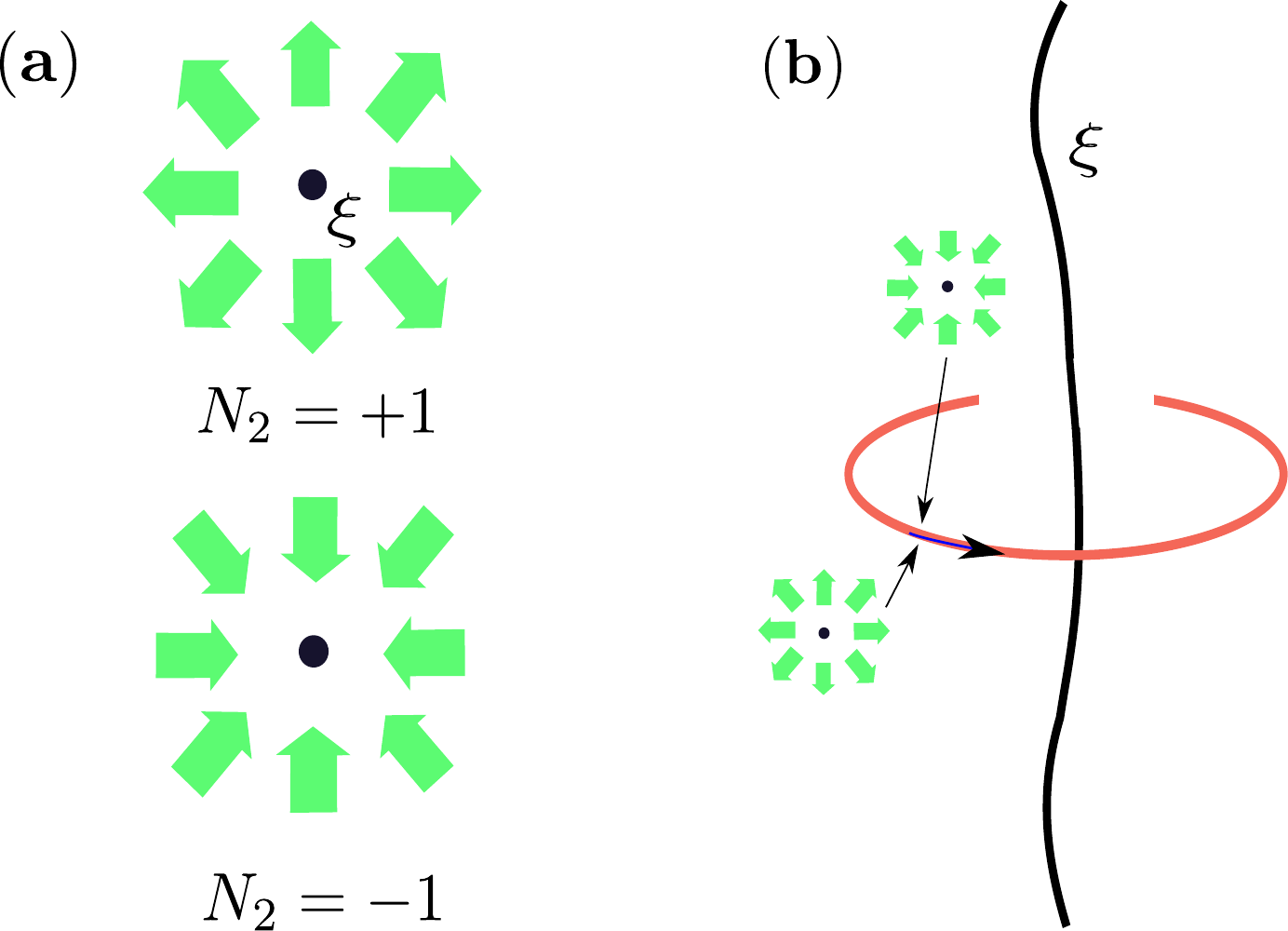}}
   \caption{Topological defects in the polar phase, which exist in addition to the  conventional vortices of group $\mathbb{Z}$ with integer circulation quanta. (a) $\hat{\mathbf{d}}$-vector monopole and anti-monopole described by the homotopy group $\pi_{2}(R_{P})$ with topological charges $N_{2}=\pm 1$ correspondingly. Their core size are on the order of the coherence length $\xi$.  The green arrows depict the configuration of $\hat{\mathbf{d}}$ vectors. (b) Half-quantum vortex (HQV) described by the $\mathbb{Z}_{2}$ subgroup of $\pi_{1}(R_{P})$ with core size $\sim \xi$. This object also got the name "Alice string", because the $\hat{\mathbf{d}}$ monopole transforms to the anti-monopole after going around the HQV,  in the same manner as it happens for the charge going around Alice string \cite{Schwarz1982}. The red circle shows the path.
   Both defects loose topological stability after transition to the PdB phase. The HQV becomes the termination line of the KLS wall bounded by string, and the monopole-hedgehog becomes the termination point of spin vortices.}
    \label{polar_defect}
    \end{figure*} 
 %%%%%%%%%%%%%
    
\section{\label{SecConventional2}Topological objects due to symmetry breaking transitions from different vacua}

In this Section we consider the topologically stable defects, which emerge at each of three symmetry breaking transitions discussed in Sec. \ref{SecConventional}.

\subsection{\label{SubNtoP}Defects in polar phase due to transition from the normal phase vacuum}

The polar phase vacuum manifold Eq. (\ref{RP}) has the homotopy groups
\begin{equation}
\pi_1(R_{\rm P}) \cong \tilde{\mathbb{Z}}\,\,, \,\, \pi_2(R_{\rm P}) \cong \mathbb{Z} \,\,, \,\, \pi_0(R_{\rm P}) \cong 0
 \,.
\label{HomotopyP}
\end{equation}
The group $\pi_1(R_{\rm P}) \cong \tilde{\mathbb{Z}}=\{n/2|n\in\mathbb{Z}\}$ includes the integers of the group $\mathbb{Z}$ via inclusion map: $n\in\mathbb{Z} \hookrightarrow n \in\tilde{\mathbb{Z}}$, which describes the conventional quantized vortices with integer winding number, and also the set of half-odd integers, i.e., $\{n+1/2|n\in\mathbb{Z}\}$. The set of half-odd integers describes vortices containing HQV, which has one-half circulation of a conventional quantized vortex. The HQVs with the topological charges $N=\pm 1/2$ are the analogs of the cosmological Alice strings. 
%In other words, the group  $\tilde{\mathbb{Z}}$ is the group $\mathbb{Z}$ multiplied by 2. (this statement is wrong)

The group $\pi_2(R_{\rm P})=\mathbb{Z}$ describes the hedgehogs (monopoles) in the $\hat{\bf d}$-field, see Fig. \ref{polar_defect}. The core size of  vortices and monopoles is on the order of the coherence length 
$\xi$. 
The topological classification of hedgehogs  is modified by the phenomenon
of influence of the homotopy group $\pi_1(R)$ on the group $\pi_2(R)$ \cite{VolovikMineev1977}. 
The monopole transforms to anti-monopole when circling around the Alice string (HQV), and thus in the presence of HQVs the hedgehogs (monopoles) of the group $\mathbb{Z}$ is reduced to the group $\mathbb{Z}_{2}$. 

In the polar phase, the HQVs have  been identified in NMR experiments \cite{Autti2016}. By applying magnetic field tilted with respect to nafen strands, one creates the soliton attached to the HQVs, which produces the measured frequency shift in the NMR spectrum. Hedgehogs (monopoles) are still not identified in superfluid $^3$He. 
\subsection{\label{Sec:defectsPdBfromN}Defects in PdB phase due to transition from normal phase vacuum}
The vacuum manifold $R_{1}$ of the PdB phase in Eq. (\ref{RPdB}) has homotopy groups
\begin{equation}
\pi_1(R_1) \cong \mathbb{Z} \times \mathbb{Z}_{2} \,\,, \,\, \pi_2(R_1) \cong 0 \,\,, \,\, \pi_0(R_1) \cong 0
 \,.
\label{HomotopyPdB}
\end{equation}
From all the defects of the polar phase with coherence length size $\xi$ in Sec. \ref{SubNtoP}, only  the  integer-quantized vortices of group $\mathbb{Z}$ survive in the PdB phase.  The other hard core defects (HQVs and hedgehogs)  are not supported by topology any more. However, the new topological object appears -- the  $\mathbb{Z}_{2}$ spin vortex, which becomes topologically stable in the PdB phase. This spin vortex is similar to that which has been observed in the bulk B-phase \cite{Kondo1992}. 
\subsection{Defects in PdB phase due to transition from polar phase vacuum}
The vacuum manifold $R_{2}$ of the PdB phase emerging at the transition from the polar phase in Eq. (\ref{PtoPDB}) has homotopy groups:
\begin{equation}
\pi_1(R_2) \cong \mathbb{Z} \,\,, \,\, \pi_2(R_2) \cong 0\,\,, \,\, \pi_0(R_2) \cong \mathbb{Z}_2
 \,.
\label{HomotopyPdBfrom Polar}
\end{equation}
These homotopy groups are responsible for  the topological defects formed in the symmetry breaking transition from the fixed degenerate vacuum of the polar phase (with $\hat{\bf d}={\rm const}$ and $\Phi={\rm const}$) to the PdB phase. 
Let us consider them separately.

\subsubsection{Spin vortices}

The homotopy group $\pi_1(R_{2}) \cong \mathbb{Z}$ describes the spin vortices with $2\pi n_1$ rotation of vectors ${\hat{\bf e}}^1$ and  ${\hat{\bf e}}^2$ about the fixed $\hat{\bf d}$ vector of the polar phase. The winding number is:
\begin{equation}
n_1 = \frac{1}{2\pi}  \oint dx^i \,\hat{\bf e}^1 \cdot \nabla_i\hat{\bf e}^2  
 \,.
\label{SpinVortexInvariant}
\end{equation}
In the vicinity of the transition, these spin vortices have the soft core of size of the coherence length, which corresponds to the transition from the polar to the PdB phase. This is $\xi/q  \gg \xi$. As distinct from the topological defects in the polar  phase, which have the "normal" core, the spin vortices in the PdB phase with $R_{2}$ have the "polar" core (quotation marks mean that in multicomponent systems the order parameter is not necessarily equal to zero on the axis of the topological defects).  Proliferation of spin vortices in PdB marks the transition to the polar phase.

As follows from Sec. \ref{Sec:defectsPdBfromN},  deep in the PdB phase only the $\mathbb{Z}_{2}$ spin vortices survive. The other spin vortices loose the topological stability and thus can live  only in the vicinity of the transition from the polar phase vacuum.  Far from transition between polar to PdB phase, their topological stability can be restored by applying the magnetic field. In this case spin vortices have the $\hat{\bf d}$ skyrmions in the core, if the winding number is even (say, doubly quantized spin vortices). The skyrmions in the $\hat{\bf d}$-field are described by the relative $\pi_2$ group and thus are the combined topological objects. All this is discussed in detail in the sections \ref{SecCombined1} and \ref{VorticesSkyrmion}.
\subsubsection{The fate of monopoles and half-quantum vortices in the PdB phase}

%%%%%%%%%%%%%%%%%%%%
\begin{figure*}%[tb!]
\centerline{\includegraphics[width=13cm,height=6cm]{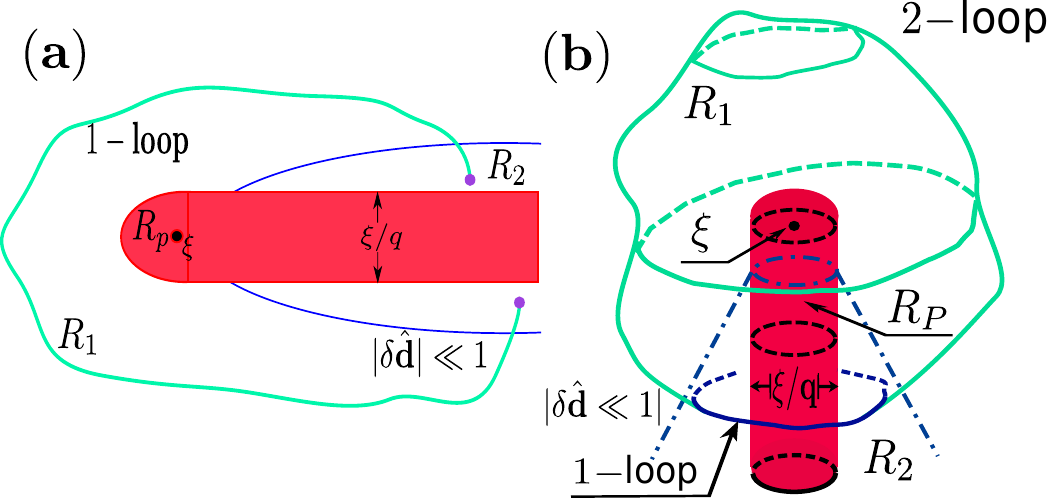}}
\caption{Illustration of topology describing the combined defects, which emerge in the two-step transition:  from the normal state to the polar phase and then from the polar phase to the polar distorted B-phase. \\
(a) KLS string wall.  In general the KLS wall is non-topological, but  it acquires the nontrivial topology in the vicinity of the second phase transition. In this limit case there are two well separated length scales: the coherence length $\xi$ of the first transition, which determines the  size of the hard core of  string (the black dot),  and the much larger coherence length $\xi/q \gg \xi$ of the second transition, which determines the soft core size of the wall (the pink region).
The hierarchy of scales gives rise to two types of the degenerate vacua in the PdB phase, $R_{1}$ and $R_{2}$. The $R_{1}$ vacua include all the degenerate vacua of the PdB phase, while the $R_2$ vacua are those, which are obtained from the fixed order parameter of the polar phase, i.e. at fixed $\hat{\mathbf{d}}$ and $\Phi$ in Eq. (\ref{AP}). This is the region, where the asymptotic condition $|\delta \hat{\mathbf{d}}| \ll 1$ is achieved.  The blue line shows the characteristic border between the regions of two classes of vacuum spaces. The topology of the string wall is determined by relative homotopy group $\pi_1(R_1,R_2)$, in which  the green loop is mapped to the space $R_1$, with the ends of the loop mapped to $R_{2}$. 
\\
(b) 
The string monopole is described by relative homotopy group $\pi_2(R_1,R_2)$. In this case the black dot shows the core of the hedgehog in the $\hat{\mathbf{d}}$-field and the pink region is the core of $4\pi$ spin vortex, which is terminated by the hedgehog. The green 2-loop is mapped to the space $R_1$, with its 1-loop edge mapped to $R_{2}$.} 
\label{R1R2loop}
\end{figure*}
%%%%%%%%%%%%%%%%%%

Since in the PdB phase $\pi_2(R_1) \cong \pi_2(R_2) \cong 0$,  the polar phase hedgehog (monopole) is topologically unstable. It becomes the termination point of  the spin vortex with two quanta (or the nexus with two singly quantized spin vortices), as discussed in detail in the section \ref{SecCombined1}. As a result the $\hat{\bf d}$-hedgehog  becomes the analog of Nambu monopole, which terminates the electroweak cosmic string \cite{Nambu1977}, see Fig. \ref{R1R2loop}(b).
The analog of  electroweak string  in the PdB phase is served either by the doubly quantized spin vortex with $n_1=2$, or by pair of $n_1=1$ spin vortices. In the presence of magnetic field, the  hedgehog (monopole) separates the string on one side of the monopole and the skyrmion on the other side of the  monopole.

\subsubsection{\label{KLSwallSec}The fate of half-quantum vortices in the PdB phase and the KLS wall}

Similar situation takes place with the HQVs, which are not topologically stable in the PdB phase. They become the termination lines of the KLS cosmic walls, as discussed in  section \ref{SecCombined1}, see Fig. \ref{R1R2loop}(a). In $^3$He experiments, after transition from the polar phase to the PdB phase in the presence of HQVs, the KLS walls appear between the neighboring vortices, and
in spite of the tension of  domain walls, the HQVs remain pinned by the nafen strands \cite{Makinen2019}.

In general the KLS wall is not topologically stable, and can be stabilized only due to symmetry reasons \cite{SalomaaVolovik1988}. 
However, in the vicinity of the transition to PdB phase from the polar phase vacuum, the KLS wall becomes topological. The topological domain wall of the thickness $\xi/q$ is described 
by the nonzero element of the homotopy group $\pi_0(R_{{\rm P}\rightarrow {\rm PdB}}) \cong \mathbb{Z}_{2}$. Example of such a wall is the domain wall between the domains with $A_{\alpha i}=\Delta_P{\rm Diag}(1, q,q)$ and $A_{\alpha i}=\Delta_P{\rm Diag}(1, q,-q)$.

\section{\label{SecCombined1}Combined objects}

\subsection{\label{SubSecCombined}Combined objects and classification by relative homotopy groups}

As mentioned before, in the vicinity of the second transition, the system has two different length scales, $\xi$ and $\xi/q \gg \xi$. This leads to the new classes of  objects, which combine the topology of both vacuum spaces, $R_{1}$ and $R_{2}$. Such combined objects are described by the relative homotopy groups \cite{MineyevVolovik1978,MMV1978,Volovik1978,GoloMonastyrky1978}
\begin{equation}
\pi_n( R_1, R_2) \,.
\end{equation}

In particular, some elements of group $\pi_1( R_1, R_2)$ describe the string wall in Fig. \ref{R1R2loop}(a) (the wall  bounded by strings, such as KLS wall \cite{Kibble1982}), while the nontrivial elements of the group $\pi_2( R_1, R_2)$ describe string monopoles in Fig. \ref{R1R2loop}(b) (such as string terminated by Nambu monopole \cite{Nambu1977}). These combined objects are topologically stable only in the vicinity of the second transition, and they loose topological stability when two length scales become comparable.

This combined topology can be illustrated by the following example of the string wall.  At small distances $\xi\ll r \ll \xi/q$  
from the core of HQV, the HQV is described by the homotopy group $\pi_1(R_{P})$. However, at larger distances $r\gg \xi/q$, 
the HQV  becomes the termination line of the wall, which is described by the $\pi_0(R_2)$ topology, see  Fig. \ref{R1R2loop}(a).  
This figure demonstrates distributions of the degeneracy spaces $R_{1}$ and $R_{2}$, which are involved in the topology of the combined object. 

The similar physics takes place for string monopoles. At small distances $\xi\ll r \ll \xi/q$  
from the core of the hedgehog, it is described by the homotopy group $\pi_2(R_{P})$, while at larger distances $r\gg \xi_q$, the monopole becomes the termination point of spin vortices described by the $\pi_1(R_2)$ topology, see Fig. \ref{R1R2loop}(b). Such combinations of $\pi_{n+1}$ and $\pi_n$ groups needed for the description of the object with two different length scales and two different dimensions are the relative homotopy groups. 
 
%%%%%%%%%%%%%%%%%%%%%%%%%
\begin{figure*}%[tb!]
\centerline{\includegraphics[width=0.7\linewidth, height=0.3\linewidth]{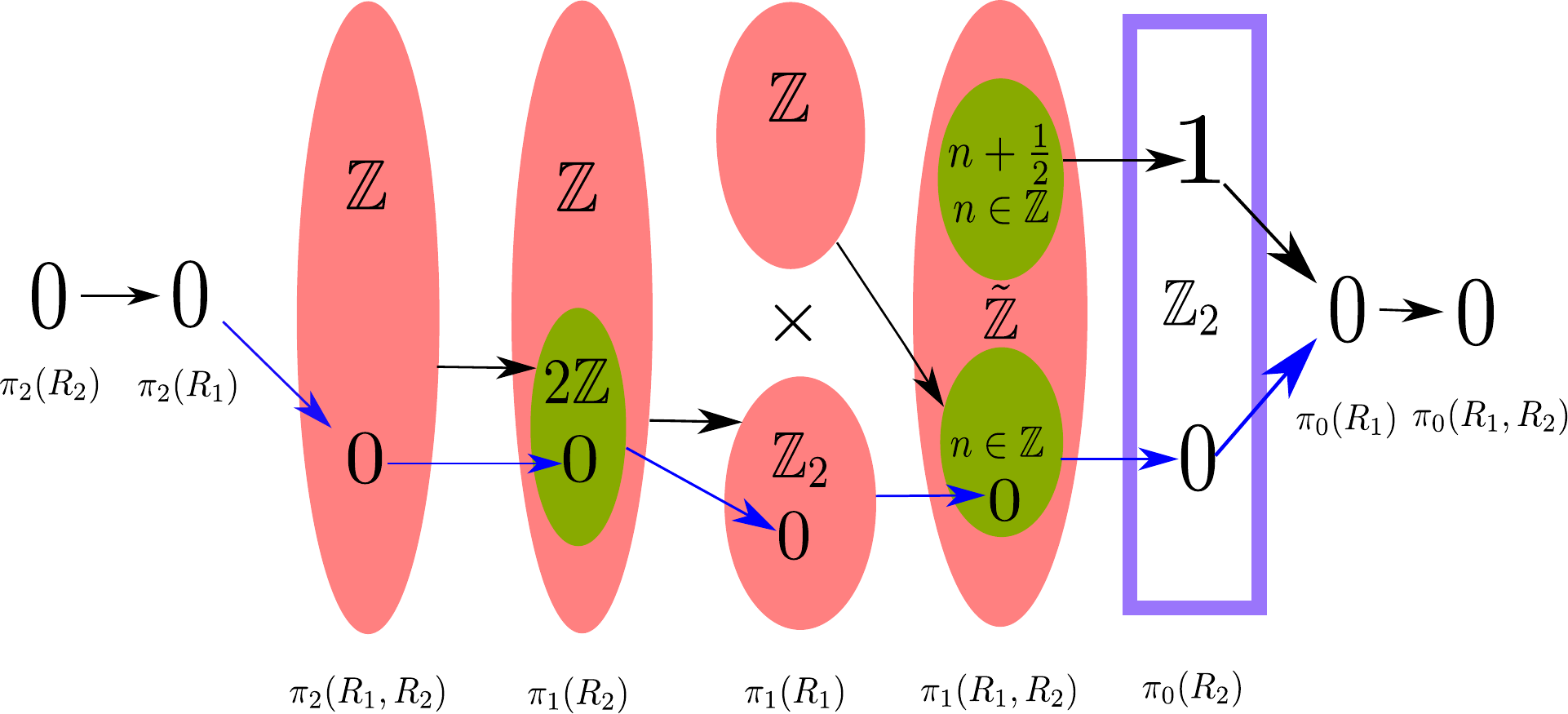}}
\caption{Illustration of the exact sequence of homomorphisms to the calculation of the
relative homotopy groups $\pi_n( R_1, R_2)$. This demonstrates that the elements of $\pi_n( R_1, R_2)$ group have two sources:  from the 
kernel of the mapping $\pi_{n-1}( R_2) \rightarrow \pi_{n-1}( R_1)$  and from the factor group of $\pi_n(R_1)$ over the image 
of the mapping $\pi_{n}( R_2) \rightarrow \pi_{n}( R_1)$. The black arrows represent the image of homomorphisms, while the blue arrows represent the kernal of every homomorphsim. This mapping diagram prescribes the relation between the elements of the composite topological defects. In particular,  it shows that  the relative homotopy group $\pi_2( R_1, R_2)$ is determined by the 
kernel of the mapping $\pi_{1}( R_2) \rightarrow \pi_{1}( R_1) =2\mathbb{Z} \cong \mathbb{Z}$. It demonstrates that 
the nontrivial monopoles are termination points of spin vortices with the total winding number being even. 
As we will see in next section, this mapping relations is identical with that describing vortex skyrmions. 
On the other hand, the relative homotopy group $\pi_{1}( R_1, R_2) \cong \tilde{\mathbb{Z}}$ is determined by both sources: 
$\mathbb{Z}_2$ which is the kernel of the homomorphism $\pi_{0}( R_2) \rightarrow \pi_{0}( R_1)$ and by the quotient group of $\pi_{1}(R_{1})$
over the image of homomorphism between $\pi_{1}(R_{2})$ and $\pi_{1}(R_{1})$. 
%related to image of homomorphisms
%$\pi_{1}( R_1)/\pi_{1}( R_2) \rightarrow \pi_{1}( R_1)\cong\mathbb{Z}$. 
As a result, there are two different kinds of phase vortices, that terminate and do not terminate the KLS wall. Those two classes of vortices consist of the two cosets of quotient $\tilde{\mathbb{Z}}/\mathbb{Z} \cong \mathbb{Z}_{2}$. Correspondingly, These are the vortices with half-odd integer circulation numbers and the vortices with integer circulation quanta.} 
\label{MappingDiagram}
\end{figure*}
%%%%%%%%%%%%%%%%%%%%%%%%

The relative homotopy groups $\pi_n( R_1, R_2)$ can be found from the following consideration. Since $R_1=G/H_{PdB}$ and $R_2=H_P/H_{PdB}$, one has
$R_1/R_2=(G/H_{PdB})(H_P/H_{PdB})=G/H_P=R_P$, and thus all the elements of the relative homotopy groups $\pi_n( R_1, R_2)$ are determined by the elements of conventional homotopy groups $\pi_n(R_P)$
 of the polar phase (we thank the Referee for this comment):
\begin{equation}
\pi_n( R_1, R_2) \cong \pi_n(R_P)
 \,.
\label{RelativeGroup}
\end{equation}
This relation demonstrates that  in the vicinity of the phase transition from the first (polar) phase to the second (PdB) phase, all the topological objects of the first phase described by the group $\pi_n(R_P)$ retain their topological charges in the second phase. Some of these defects  remain free, while the other become the parts of the composite defects --  the string monopole and for the KLS string wall in Secs. IV B and IV C respectively.

The Eq.(\ref{RelativeGroup}) does not resolve between the free and the composite objects of the second phase.  The full classification of the topological
objects in the second phase depends not only on $R_1/R_2$, but also on the details of the mappings in
 the exact sequence of homomorphisms, which is calculated in
appendices Sec. \ref{app:derivation} and \ref{RHG}. 
The mapping diagram of exact sequence in Fig. \ref{MappingDiagram} depicts the relation between different topological objects in $R_{1}$ and $R_{2}$. 

\subsection{Wall bounded by string -- KLS string wall}

%%%%%%%%%%%%%%%%%%%%%%%%%%
\begin{figure*}%[tb!]
\centerline{\includegraphics[width=0.7\linewidth, height=0.4\linewidth]{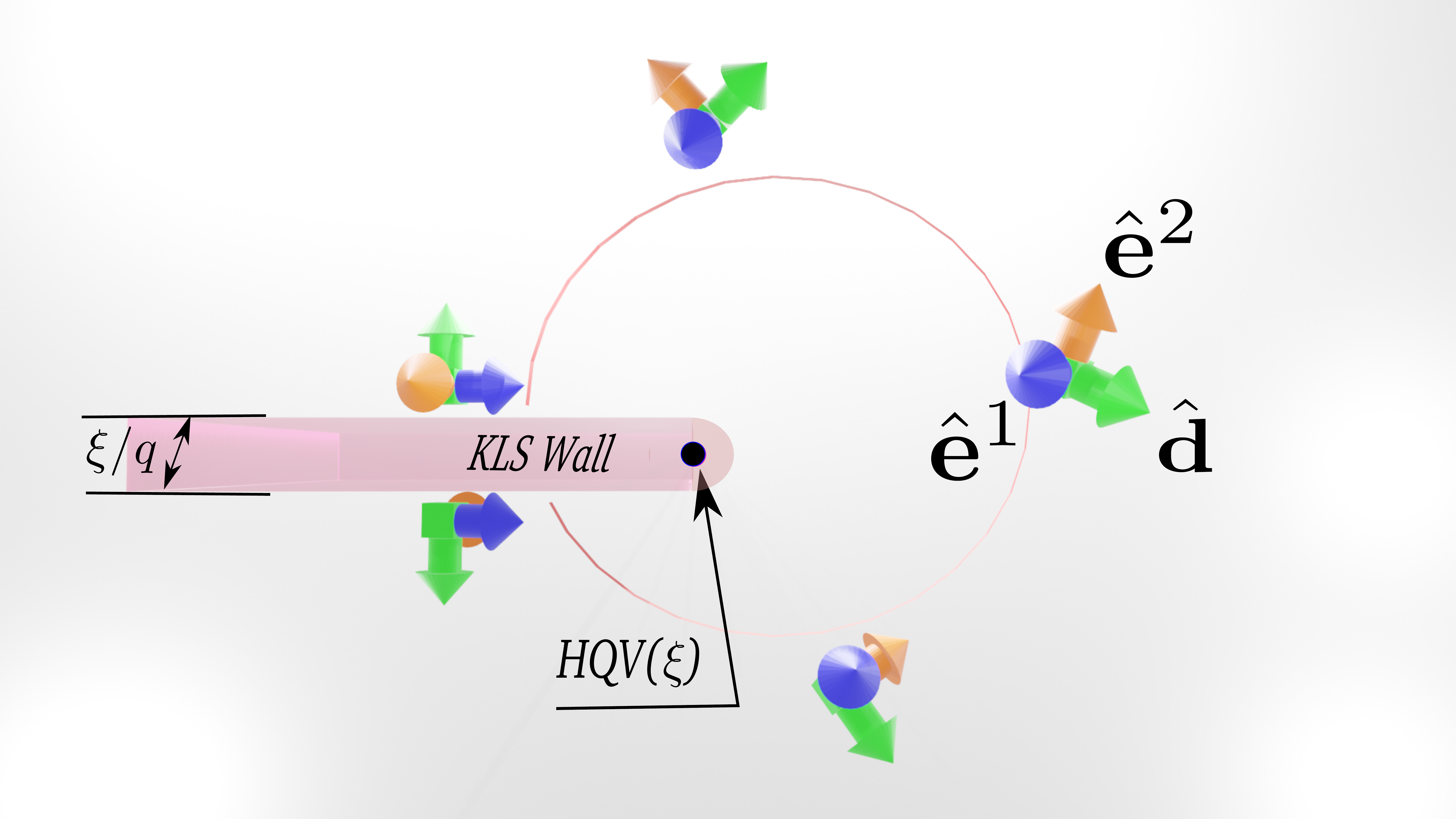}}
\caption{Illustration of the KLS string wall.
The wall is topologically protected if the order parameter takes values from disconnected parts of $R_{2}$. In this case it is the combined object -- the HQV, which terminates the KLS wall.  The pink region is the topological KLS wall with thickness $\xi/q$, while the black small dot is HQV string, which diameter is $\xi$. The spin tripods show the configurations of order parameter around the HQVs string. The flipping of the tripods on two sides of KLS wall demonstrates the Domain wall feature.  
} 
\label{HoleInWallFig}
\end{figure*}
%%%%%%%%%%%%%%%%%%%%%%%%

The relative homotopy group is
\begin{equation}
\pi_1( R_1, R_2) \cong \pi_1(R_P) \cong \tilde{\mathbb{Z}} \,.
\label{pi1relative}
\end{equation}
This shows that the topological charges of vortices in the second (PdB) phase are the same as in the first (polar) phase. In both phases they form the group  of integer and half-odd integers $N$. However, the physical realizations of these vortices are different in the two phases. Vortices with integer $N$ remain free, while
 vortices with half-odd integers, $N=k+1/2$, terminate the wall bounded by string --  the KLS string wall. Fig. \ref{HoleInWallFig} illustrates the configuration of the composite object. 
This kind of topologically protected KLS string wall induces the cosmological catastrophe in the axion solution of strong \textit{CP} problem \cite{Sikivie1982,Andrea2019,Chatterjee2019}. 

In general, in the vicinity of the second phase transition the topological objects of the first phase remain topological in the second phase. Some of these defects remain free, while the other become the part of the composite objects with the same topological invariants. The separation between these two groups of the objects is determined by the mappings in
 the exact sequence of homomorphisms.

\subsection{Strings terminated by monopole -- String Monopole}

\begin{figure*}%[tb!]
\centerline{\includegraphics[width=0.7\linewidth, height=0.35\linewidth]{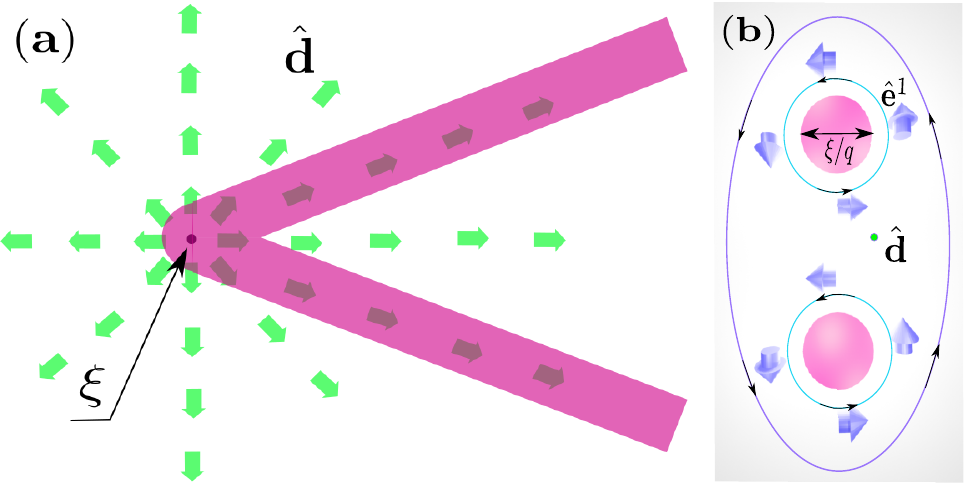}}
\caption{Illustration of string monopole in the PdB phase.  The monopole in the  $\hat{\bf d}$-field has topological charge $n_{2}=1$ and hard core of the coherence length size $\xi$ (black dot). This string monopole terminates two spin vortices  with the soft core size $\xi/q$ and with the total topological charge $n_{1}=1+1=2=2n_{2}$ according to Eq. (\ref{MonopoleInvariant}). (a) Texture of $\hat{\mathbf{d}}$ vector string monopole with $n_{2}=2$ terminating two spin vortices.  The green arrow corresponds to $\hat{\mathbf{d}}$ vector. The red region is the soft core of spin vortex of size $\xi/q$. (b) The cross sections of  two spin vortices with $n_{1}=1$ each. In these cross sections, 
$\hat{\mathbf{e}}_{1}$ and $\hat{\mathbf{e}}_{2}$ vectors rotate around $\hat{\mathbf{d}}$ vector by $2\pi$ for every spin vortex. The core size of every spin vortex is $\xi/q$. The blue line is the field line of total $\hat{\mathbf{e}}$ vectors rotation, while the green lines are 
$\hat{\mathbf{e}}$ vectors field lines of every spin vortex.} 
\label{MonopoleFig1}
\end{figure*}
%%%%%%%%%%%%%%%%%%%%%%%%

The relative homotopy group 
\begin{equation}
 \pi_{2}(R_{1},R_{2})\cong\pi_2(R_P)\cong \mathbb{Z} 
 \,,
\label{pi2relative}
\end{equation}
describes monopoles (hedgehogs) of $\hat{\mathbf{d}}$ field. They survive in  the vicinity of the second transition as the topological objects which terminate the spin vortices. The corresponding composite object -- the  string monopole -- has two topological charges, $n_{2}$ and $n_{1}$, which are related as: 
\begin{equation}
n_2= \frac{1}{8\pi} e^{ijk} \int_{\rm S^{2}} dS_k \,\hat{\bf d}\cdot 
\left(  \nabla_i \hat{\bf d} \times \nabla_j \hat{\bf d} \right)=\frac{1}{2}n_1
 \,.
\label{MonopoleInvariant}
\end{equation}
Here $S^{2}$ is the surface encircling monopole and the group $\mathbb{Z}$  is the group of integers $n_2$ --  the topological charges of the hedgehog. The $n_{1}$ is the winding number of spin vortices in Eq. (\ref{SpinVortexInvariant}). The equation $n_1=2n_2$ in (\ref{MonopoleInvariant}) shows  that monopole can be termination point of spin vortices with the even total charge $n_2$. This situation is similar to the monopole  in the chiral A-phase \cite{volovik2000,Blaha1976,Volovik1976}, which either terminates a single vortex with $n_{1}=2$, or forms the nexus with two singly quantized vortices with $n_{1}=1+1=2$, or with four HQVs with $n_{1}=1/2+1/2+1/2+1/2=2$. Those vortices, which connect with monopoles ($n_{2}>0$) or antimonopoles ($n_{2}<0$) allow the existences of complex monopole-antimonopole networks \cite{Kibble2015,Saurabh2019,Shafi2019,Volovik2019d}.

Fig. \ref{MonopoleFig1} illustrates the configuration of the string monopole, which consists of the hedgehog with $n_{2}=1$ and two strings -- spin vortices each with $n_1=1$. The spin vortices have a soft core with size $\xi/q$.

\section{\label{VorticesSkyrmion}Skyrmions and Nexus in the presence of magnetic field}

%%%%%%%%%%%%%%%%%%%%%%%%%%%%
\begin{figure*}%[tb!]
\centerline{\includegraphics[width=0.65\linewidth]{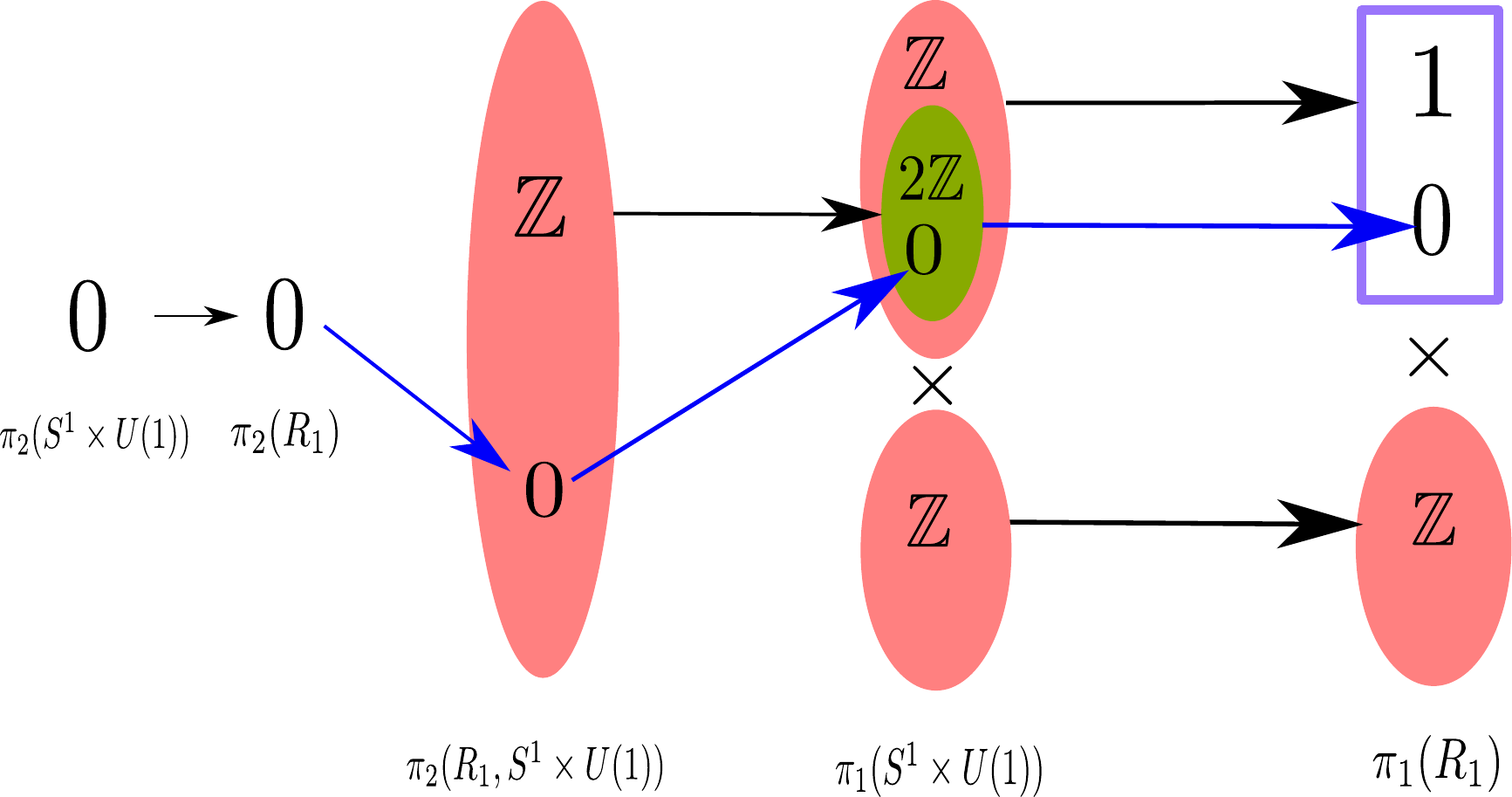}}
\caption{Mapping diagram of exact sequence between $R_{1}$ and $S^{1} \times U(1)$. The black arrows represent the image of homomorphisms, while the blue arrows represent the kernal of every homomorphsim. This diagram shows 
that the skyrmions soften the core of $\mathbb{Z}$ spin vortices with size $\xi/q$ to size $\xi_{H}$ in the presence of magnetic field. In regions larger than $\xi_{\mathbf{H}}$, vortex skyrmions can be connected with spin vortices via string monopole, if their total topological charge is even according to Eq. (\ref{SkyrmInvariant}). This is because $\pi_{2}(R_{1},R_{2}) \cong \pi_{2}(R_{1},S^{1} \times U(1))$. There are also phase vortices described by $\mathbb{Z}$, but here we ignore them because they do not influence the connection between the spin vortices and skyrmions. } 
\label{MappingDiagram_SN}
\end{figure*}
%%%%%%%%%%%%%%%%%%%%%%%%%%%%

In the presence of magnetic field $\mathbf{H}$, a new length scale appears in the PdB phase -- the magnetic length 
$\xi_{H} \propto |\mathbf{H}|^{-1}$.  The magnetic length  $\xi_{H}$ is the longest length scale if we neglect the spin-orbit coupling. In this case, one obtains the two scale system of type (i) in Introduction Section. In the region with length scale larger than $\xi_{H}$, the magnetic anisotropy locks the directions of $\hat{\mathbf{d}}$ vector in the  plane perpendicular to $\mathbf{H}$ to minimize the magnetic energy, which is proportional to $|\mathbf{H} \cdot \hat{\mathbf{d}}|^{2}$.  The degenerate space of the  order parameter is reduced from $R_1 \cong SO_{L-S}(3) \times U(1)$ in Eq. (\ref{RPdB}) to 
$R_1^H =S^{1} \times S^{1} \times U(1)$ in the regions which are larger than $\xi_{H}$. The first $S^{1}$ is the manifold of in plane $\hat{\mathbf{d}}$ vector, while the second $S^{1}$ is the manifold of rotations of $\hat{\mathbf{e}}^{1}$ and $\hat{\mathbf{e}}^{2}$ about the $\hat{\mathbf{d}}$-axis. Then  the second relative homotopy group of combined objects with length scale $\xi_{H}$ is $\pi_{2}(R_{1},R_1^H)\cong \mathbb{Z}\times \mathbb{Z}$. However, for $q\ll 1$, the gradient energy of the $\hat{\mathbf{d}}$-textures is much larger than that of the textures in $\hat{\mathbf{e}}^{1}$ and $\hat{\mathbf{e}}^{2}$ fields \cite{VollhardtWolfle1990}. That is why we consider only the $S^{1}$ manifold of $\hat{\mathbf{e}}^{1}$ and $\hat{\mathbf{e}}^{2}$, and neglect the $S^{1}$ manifold of  $\hat{\mathbf{d}}$. Then the relative second homotopy group which we need in this case is 
\begin{equation}
\pi_{2}(R_{1},S^{1} \times U(1)) \cong \mathbb{Z}.
\label{RHG_SkyrmionNexus}
\end{equation}
These results for the relative homotopy group have been confirmed by calculations using the exact sequence, see details in appendices Sec. \ref{app:derivation} and Sec. \ref{RHG}.
The mapping diagram of exact sequence is shown in Fig. \ref{MappingDiagram_SN}. 

%%%%%%%%%%%%%%%%%%%%%%%%%%%%%%% 
\begin{figure*}%[tb!]
\centerline{\includegraphics[width=0.65\linewidth, height=0.3\linewidth]{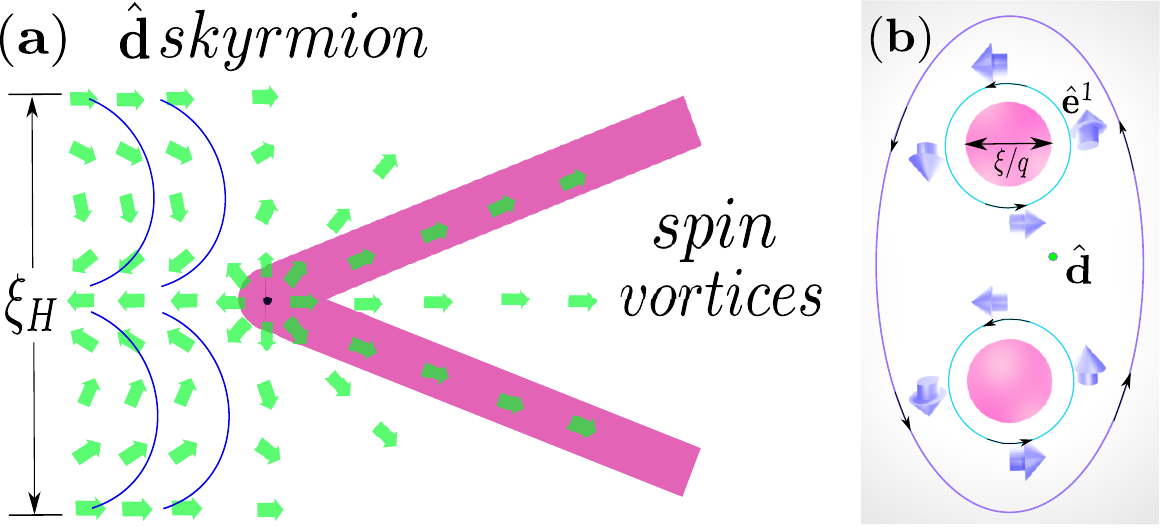}}
\caption{Illustration of nexus object in presence of magnetic field. The nexus connects spin vortices with core size $\xi/q$ and the vortex skyrmion with core size $\xi_H$, where $\xi_{H} \gg \xi/q$. (a) the texture configuration of nexus object with $n_{2}=1$ and $n_{1}=2$. The green arrows are the $\hat{\mathbf{d}}$ vectors and the pink regions are core regions of defects. The black dot is the core of string monopole with size $\xi$, while the pink regions are the cores of spin vortices with size $\xi/q$. The green arrows represent the distribution of $\hat{\mathbf{d}}$ vectors. The vortex skyrmion with core size $\xi_{H}$ transforms to two spin vortices via nexus.  (b) the cross section of two spin vortices. Every spin vortex has the $2\pi$ rotation of $\hat{\mathbf{e}}^{1}$ and  $\hat{\mathbf{e}}^{2}$ vectors around fixed $\hat{\mathbf{d}}$ vector. The blue arrows represent the field of the $\hat{\mathbf{e}}^{1}$ vector. }
\label{NexusFig}
\end{figure*}
%%%%%%%%%%%%%%%%%%%%%%%%%%%%%%%

The relative homotopy group $\pi_{2}(R_{1},S^{1} \times U(1))$ describes the composite object in the PdB phase in the presence of magnetic field. This object is the spin vortex with even winding number, which has the soft core of size $\xi_{H}$ represented as skyrmion, see Fig. \ref{NexusFig}. The topological charge of skyrmion is 
\begin{equation}
n_2= \frac{1}{8\pi} e^{ijk} \int_{\rm D_{2}} dS_k \,\hat{\bf d}\cdot 
\left(  \nabla_i \hat{\bf d} \times \nabla_j \hat{\bf d} \right)=\frac{1}{2}n_1
 \,,
\label{SkyrmInvariant}
\end{equation}
where $D_{2}$ is the cross-section of skyrmion and $n_{1}$ is the winding number of spin vortices in Eq. (\ref{SpinVortexInvariant}). 
The Eq. (\ref{SkyrmInvariant}) is the analog of the Mermin-Ho relation in $^3$He-A \cite{MerminHo1976}.
Eq. (\ref{SkyrmInvariant}) is identical with Eq. (\ref{MonopoleInvariant}) because of $\pi_{2}(R_{1},S^{1} \times U(1)) \cong \pi_{2}(R_{1},R_{2})$. Due to this relation the vortex skyrmion can be connected to $\mathbb{Z}$ spin vortices with core size $\xi/q$ via the string monopole. 
Such composite objects, where the monopole connects several linear objects is called  nexus.  It demonstrates the interplay between $\pi_1$ and $\pi_2$ topologies. 

Originally vortex skyrmions formed by orbital and phase degenerate parameters have been suggested in $^3$He-A by Anderson and Toulouse \cite{Anderson1977} 
and by Chechetkin \cite{Chechetkin1976}. The lattice of vortex-skyrmions in rotating $^3$He-A has been discussed in Ref. \cite{Volovik1977}.
 These objects have been identified in different experiments made under rotation \cite{Seppala1984,Pekola1990}. The dynamics of the vortex skyrmions  provides an effective electromagnetic fields, which induces the observed effect of chiral anomaly experienced by fermionic excitations (Weyl fermions)  living in the soft core of a vortex skyrmion \cite{Bevan1997}. 

\section{conclusion and discussion}

Here we discussed the topology, which emerges in two-step phase transition in the vicinity of the second transition.
An example is provided by the second order phase transition from the normal $^3$He to the polar phase followed by the second order phase transition from the polar phase to the PdB phase
experimentally observed in superfluid $^3$He in nafen \cite{Makinen2019}. Here the composite object -- the analog of the KLS wall bounded by cosmic string has been observed. We demonstrated that in the vicinity of the second transition, such composite object is described by the relative homotopy groups. The reason for that is the existence of the two well separated length scales. The smaller length scale determines the core size of the half-quantum vortex (the analog of Alice cosmic string). It is the coherence length $\xi$ related to the symmetry breaking phase transition form the normal liquid to the polar phase. The larger length scale $\xi/q\gg \xi$ determines the soft core size of the KLS wall terminated by this string. It is the coherence length related to the second symmetry breaking phase transition -- the  transition form the polar phase to the PdB phase.

The two-scale composite defects are described by relative homotopy groups $\pi_n(R_1,R_2)$. Here $R_1$ is the vacuum manifold of the PdB phase, while $R_2$ is also the vacuum manifold of the PdB phase, but at a fixed value of the order parameter of the polar phase before the transition to the PdB phase. The observed  KLS wall terminated by the half-quantum vortex is determined 
by the nontrivial element of $\pi_1(R_1,R_2)$. The other composite object, which is still waiting for its observation, is the monopole (hedgehog), which terminates the string (the spin vortex). Its topology is determined 
by the nontrivial element of $\pi_2(R_1,R_2)$. The core of the monopole is of coherence length size $\xi$, while the spin vortices have the soft core of size $\xi/q\gg \xi$. The relative homotopy groups $\pi_n(R_1,R_2)$ are calculated using  the exact sequence of the group homomorphisms. 

The topology of these combined objects demonstrates new application of the  relative homotopy groups. Earlier the relative homotopy groups have been applied for classification of topological defects on the surface of the ordered system \cite{Volovik1978}, and for classification of topological solitons terminated by point or linear defects \cite{MineyevVolovik1978}. 

We also considered the more complicated object -- the nexus, which combines the monopole, the string terminated by monopole, and skyrmion (topological soliton) terminated by the same monopole. Such object in the PdB phase arises in the presence of  magnetic field, which provides another length scale. The situation becomes even richer, when the spin-orbit interaction is included, which provides the fourth length scale and extends the multi-scale topology. The  objects combining vortices and skyrmions were recently considered for superconductor-ferromagnet heterostructures, in which the existence of Majorana bound states were suggested \cite{Samme2019,Stefan2019}.

Typically the state of the system  with  topological defects represents the excited state of the system. However, the topological defects can form  the ground state. Earlier it was suggested that the suppression of the B-phase on the boundary of superfluid $^3$He may lead to formation of stripe phase in superfluid $^3$He-B under nanoscale confinement in a slab geometry \cite{Vorontsov2007}. On a microscopic level, this inhomogeneous phase is thought as the periodic array of the KLS domain walls between the degenerate states of the B-phase, see Refs. \cite{Volovik1990,SalomaaVolovik1988}. 
 The possible observation of such spatially modulated phase has been reported \cite{Shook2019,Levitin2019}. 
 Similar situation may take place in another kind of confined geometry, in nafen. The strands of nafen could play the same role as the boundaries in the slab confinement. The suppression of the order parameter near the strands may result in the spontaneous proliferation of the composite defects leading to the stripe phases or stripe glasses. 
 
\begin{acknowledgments}
We thank Jaakko Nissinen and Ilya M. Eremin for discussions.  This work has been supported by the European
Research  Council  (ERC)  under  the  European  Union’s
Horizon 2020 research and innovation programme (Grant Agreement No.  694248).
\end{acknowledgments}

\appendix
\section{\label{app:derivation}Exact Sequences: with and without magnetic field}

\subsubsection{No Magnetic Field} 

The exact sequence of (relative) homotopy groups means that the image of any homomorphism $x_{*}^{n}:A \rightarrow B$ (the sets of the elements of the group $B$ into which the elements of $A$ are mapped) is the kernel of the next homomorphism $x_{*}^{n+1}:B \rightarrow C$ (the sets of the elements of $B$ which are mapped to the zero element of $C$) i.e. $\operatorname{im}x^{n}_{*} \cong {\ker}x^{n+1}_{*}$, with $n\in \mathbb{Z}$ \cite{Nash1988}. The relative homotopy classes of $\pi_{k+1}(R_{1},R_{2})$ are mapped to the homotopy classes of $\pi_{k}(R_{2})$ by mapping the $k$-dimension subset of $k+1$ sphere, which surrounds the defects, into $R_{2}$. This mapping between two homotopy classes with different dimensions is called boundary homomorphism $\partial_{*}$ \cite{Nash1988}. Boundary homormophism shows how topological objects with different dimensions connect to each other. In the PdB phase the exact sequence of homomorphisms is
\begin{widetext}
\begin{equation}
\xymatrix@1@R=10pt@C=12pt{
\pi_2( R_2) \ar@{-}[d] \ar[r]^{i_{*}} & \pi_2(R_1) \ar@{-}[d] \ar[r]^{j_{*}} &
\pi_2( R_1, R_2) \ar@{-}[d] \ar[r]^-{{\partial}^{k}_{*}} & \pi_1(R_2) \ar@{-}[d] \ar[r]^{m_{*}} &
\pi_1( R_1) \ar@{-}[d] \ar[r]^{n_{*}} & \pi_1(R_1,R_2) \ar@{-}[d] \ar[r]^-{{\partial}^{p}_{*}} & \pi_0( R_2) \ar@{-}[d] \ar[r]^{q_{*}} & \pi_0(R_1) \ar@{-}[d] \ar[r]^-{r_{*}} & \pi_0(R_1,R_2) \ar@{-}[d] \\
0 \ar[r]^{i_{*}} & 0 \ar[r]^{j_{*}} & \mathbb{Z} \ar[r]^{{\partial}^{k}_{*}} &
\mathbb{Z} \ar[r]^{m_{*}} & \mathbb{Z} \times \mathbb{Z}_{2} \ar[r]^{n_{*}} & \tilde{\mathbb{Z}} \ar[r]^{{\partial}^{p}_{*}} & \mathbb{Z}_{2} \ar[r]^{q_{*}} & 0 \ar[r]^{r_{*}} & 0 
}
\label{sequence}
\end{equation}
\end{widetext}
where the $\partial^{k}_{*}$ and $\partial^{p}_{*}$ are boundary homomorphisms. This gives the following relative homotopy groups:
%\begin{widetext}
%\begin{equation}
%\pi_2( R_1, R_2) \cong \mathbb{Z}^{+} \cong \mathbb{Z}/\mathbb{Z}_{2} \cong \mathbb{Z} \,\,,\,\,\pi_1( R_1, R_2) \cong \mathbb{Z}_{2}\times \mathbb{Z} \,\,,\,\,\pi_0( R_1, R_2) \cong 0
% \,,
%\label{RelativeSequence}
%\end{equation}
%\end{widetext}
$\pi_2( R_1, R_2) \cong \mathbb{Z}$, $\pi_1( R_1, R_2) \cong  \tilde{\mathbb{Z}}$ and $\pi_0( R_1, R_2) \cong 0$. The $\partial^{k}_{*}$ maps the homotopy classes of string monopoles to the homotopy classes of spin vortices. The $\partial^{p}_{*}$ maps the homotopy classes of KLS string wall to homotopy classes of domain wall. The kernels and images of every relative homotopy group are analyzed in the section \ref{RHG} of appendices. 

\subsubsection{In the presence of magnetic field}

In the presence of magnetic field the corresponding exact sequence is 
\begin{widetext}
\begin{equation}
\xymatrix@1@R=10pt@C=13pt{
\pi_2(S^{1} \times U(1)) \ar@{-}[d] \ar[r]^-{i_{*}} & \pi_2(R_1) \ar@{-}[d] \ar[r]^-{j_{*}} &
\pi_2( R_1, S^{1} \times U(1)) \ar@{-}[d] \ar[r]^-{{\partial}^{k}_{*}} & \pi_1(S^{1} \times U(1)) \ar@{-}[d] \ar[r]^-{m_{*}} &
\pi_1( R_1) \ar@{-}[d]\\
0 \ar[r]^{i_{*}} & 0 \ar[r]^{j_{*}} & \mathbb{Z} \ar[r]^{{\partial}^{k}_{*}} &
\mathbb{Z}\times \mathbb{Z} \ar[r]^{m_{*}} & \mathbb{Z}_{2} \times \mathbb{Z}  \\
}
\label{sequence_SkyrmionNexus}
\end{equation}
\end{widetext}
i.e. $\pi_{2}(R_{1},S^{1} \times U(1)) = 2\mathbb{Z} \cong \mathbb{Z}$. We found $\ker \partial^{k}_{*} \cong 0$ and $\operatorname{im} \partial^{k}_{*} = 2\mathbb{Z} \cong \mathbb{Z}$. That means that  only those objects are topological protected, which have an even total winding number of spin rotation. These objects are the $\hat{\mathbf{d}}$-vector skyrmions. Since $\pi_{2}(R_{1},S^{1} \times U(1)) \cong \pi_{2}(R_{2},R_{1})$, these $\hat{\mathbf{d}}$-skyrmions can terminate on the $\hat{\mathbf{d}}$-monopole, which in turn is the end point of spin vortices with the total even number of spin rotation. As a result one obtains the composite effect -- the nexus in Fig. \ref{NexusFig}.  

\section{\label{RHG}relative homotopy groups}

\subsection{$\pi_{2}(R_{1},R_{2})$}

The objects described by the $\pi_{2}(R_{1},R_{2})$ are monopoles of $\hat{\mathbf{d}}$-vector, because the manifold of the degenerate states of the $\hat{\mathbf{d}}$-vector is $S^{2} \cong SO_{S}(3)/SO_{S}(2)$, and we have the mapping from $S^{2}$ in real space to $S^{2}$ manifold of $\hat{\mathbf{d}}$-vectors.

The boundary homomorphism $\partial^{k}_{*}$ maps $S^{1} \subset S^{2}$ to $R_{2}$. Then the $\operatorname{im} \partial^{k}_{*}$ describes all classes of string defects terminated by the monopoles. We found $\operatorname{im} {\partial_{*}^{k}} = 2\mathbb{Z}\cong \mathbb{Z}$ which is the set of even numbers. This means that only the spin vortices with the total even winding number can form the string monopole. This situation is similar to the monopole connected with four half-quantum vortices in the A-phase, where the total winding number is 2 \cite{volovik2000}.
The topologically trivial monopole cannot connect with the string defects because of ${\ker}{\partial^{k}_{*}} \cong 0$. Actually this trivial class is identical to $\pi_{2}(R_{1})$ because ${\ker}{j_{*}} \cong \operatorname{im} {i_{*}} \cong 0$. 

\subsection{$\pi_{1}(R_{1},R_{2})$}

The relative homotopy group is $\pi_1( R_1, R_2) \cong \tilde{\mathbb{Z}}$. From ${\ker}q_{*} \cong \operatorname{im}\partial^{p}_{*} \cong \pi_{0}(R_{2}) \cong \mathbb{Z}_{2} $, we know there are domain walls bounded by string defects. The set of half-odd integers of the group $\tilde{\mathbb{Z}}$, which come from $\operatorname{im} \partial^{p}_{*}$ describes the domain wall terminated by string defects -- 
 the KLS wall terminated by HQV or by any vortex with half-odd integer winding number $N=k+1/2$. The vortices, which come from ${\ker}\partial^{p}_{*}\cong\mathbb{Z}$ are vortices with integer winding number. These vortices are free. 
   
\subsection{$\pi_{1}(R_{1},S^{1} \times U(1))$}

From exact sequence in Eq. (\ref{sequence_SkyrmionNexus}) it follows that $\pi_{2}(R_{1},S^{1} \times U(1)) = 2\mathbb{Z}_{2} \cong \mathbb{Z}$. 
This group describes the linear skyrmions in the $\hat{\mathbf{d}}$-vector, which are the linear analogs of the original point-like skyrmion \cite{skyrme1962,khawaja2001}. 
The spin texture inside the cross-section $D_{2}$ of the skyrmion corresponds to continuous mapping to $SO_{S}(3)$, which is implemented by choosing first a direction of $\hat{\mathbf{d}}$  and then making $SO_{S}(2)$ rotation of $\hat{\mathbf{e}}^{1}$ and  $\hat{\mathbf{e}}^{2}$ around this direction.  This skyrmion also represents  the spin vortex with even winding number, because of $\operatorname{im} \partial_{*}^{k} \cong 2\mathbb{Z}$ and ${\ker}\partial_{*}^{k} \cong 0$.


\begin{thebibliography}{99}

\bibitem{Nambu1977}
Y. Nambu, 
String-like configurations in the Weinberg-Salam theory,
Nucl. Phys. B {\bf 130}, 505 (1977).

\bibitem{Kibble1982}
T. W. B. Kibble, G. Lazarides, and Q. Shafi,
Walls bounded by strings,
Phys. Rev. D {\bf 26}, 435--439 (1982);
Strings in $SO(10)$,
Phys. Lett. B {\bf 113}, 237--239 (1982).

\bibitem{Kibble2000}
T. W. B. Kibble,
 Classification of Topological Defects and Their Relevance to Cosmology and Elsewhere,
 In: Bunkov Y.M., Godfrin H. (eds) Topological Defects and the Non-Equilibrium Dynamics of Symmetry Breaking Phase Transitions, NATO Science Series (Series C: Mathematical and Physical Sciences), vol. {\bf 549}, pp. 7--31, Springer, Dordrecht  (2000).

\bibitem{Kibble2015}
T. W. B. Kibble and T. Vachaspati,
Monopoles on strings,
J. Phys. G {\bf 42}, 094002 (2015).

\bibitem{Vilenkin1982}
A. Vilenkin and A.E. Everett, 
Cosmic strings and domain walls in models with Goldstone and pseudo-Goldstone bosons, 
Phys. Rev. Lett. {\bf 48}, 1867 (1982).

\bibitem{Sikivie1982}
P. Sikivie, Of Axion, 
Domain walls and the early Universe, 
Phys. Rev. Lett. {\bf 48}, 1156 (1982).

\bibitem{Andrea2019}
A. Caputo, M. Reig,
Cosmic implications of a low-scale solution to the axion domain wall problem, 
Phys. Rev. D {\bf 100}, 063530 (2019). 

\bibitem{Chatterjee2019}
C. Chatterjee, T. Higaki, and M. Nitta,
Note on a solution to domain wall problem with the Lazarides-Shafi
mechanism in axion dark matter models,
Phys. Rev. D {\bf 101}, 075026 (2020). 

\bibitem{Makinen2019}
 J.T. M\"akinen, V.V. Dmitriev, J. Nissinen, J. Rysti, G.E. Volovik, A.N. Yudin, K. Zhang, V.B. Eltsov,
Half-quantum vortices and walls bounded by strings in the polar-distorted phases of topological superfluid $^3$He,
Nat. Comm. {\bf 10}, 237 (2019).

\bibitem{Kazushi2006}
K. Aoyama, R. Ikeda, 
Pairing states of superfluid $^3$He in uniaxially anisotropic aerogel, 
Phys. Rev. B {\bf 73}, 060504 (2006).

\bibitem{Ikeda2014}
S. Yang, R. Ikeda,
Possibility of unconventional pairing states in superfluid $^3$He in uniaxially anisotropic aerogels,
J. Phys. Soc. Jpn {\bf 83}, 084602 (2014).

\bibitem{Askhadullin2012} 
R.Sh. Askhadullin, V.V. Dmitriev, D.A. Krasnikhin, P.N. Martynov, A.A. Osipov, A.A. Senin, A.N. Yudin, 
Phase diagram of superfluid 3He in "nematically ordered" aerogel,
JETP. Lett. {\bf 95}, 326 (2012).


\bibitem{Anderson1959}
 P. W. Anderson, 
Theory of dirty superconductors,
J. Phys. Chem. Solids {\bf 11}, 26--30 (1959).

\bibitem{Fomin2018}
I.A. Fomin,
Analog of Anderson theorem for the polar phase of liquid $^3$He in nematic aerogel,
JETP {\bf 127}, 933--938 (2018).

\bibitem{Fomin2020}
I.A. Fomin,
Temperature dependence of the order parameter of the polar phase of liquid $^3$He in nematic aerogel,
arXiv:2003.09652.

\bibitem{Ikeda2019b}
M. Tange, R. Ikeda,
Half-quantum vortex pair in polar-distorted B phase of superfluid $^3$He in aerogels,
Phys. Rev. B. \textbf{101}, 094512 (2020).

\bibitem{Ramires2018}
A. Ramires, D. F. Agterberg and M. Sigrist, 
Tailoring  $T_c$  by symmetry principles: The concept of superconducting fitness,
Phys. Rev. B {\bf 98}, 024501 (2018).

\bibitem{Eltsov2019}
V.B.  Eltsov,T.  Kamppinen,J. Rysti, and  G.E.  Volovik, 
Topological nodal line in superfluid $^3$He and the Anderson theorem, 
arXiv:1908.01645 (2019).

\bibitem{Autti2016}
S. Autti, V.V. Dmitriev, J.T. M\"akinen, A.A. Soldatov, G.E. Volovik,
A.N. Yudin, V.V. Zavjalov, and V.B. Eltsov,
Observation of half-quantum vortices in superfluid $^3$He,
Phys. Rev. Lett. {\bf 117}, 255301 (2016)


\bibitem{VolovikMineev1976} 
G.E. Volovik and V.P. Mineev, 
Line and point singularities in superfluid $^3$He,
JETP Lett. {\bf 24}, 561--563 (1976).

\bibitem{Cross1977}
M.C. Cross and W.F.  Brinkman,
Textural singularities in superfluid A-phase of $^3$He,
J. Low Temp. Phys. {\bf 27}, 683--686 (1977).
 
\bibitem{Salomaa1985}
M.M. Salomaa, G.E. Volovik,
Half-quantum  vortices in superfluid $^3$He-A,
Phys. Rev. Lett. {\bf 55}, 1184--1187 (1985).

\bibitem{Hu1987}
Chia-Ren Hu and K. Maki,
Satellite magnetic resonances of a bound pair of half-quantum vortices in rotating superfluid  $^3$He-A,
Phys. Rev. B {\bf 36}, 6871--6880 (1987).

\bibitem{Vakaryuk2009}
V. Vakaryuk and A.J. Leggett,
Spin polarization of half-quantum vortex in systems with equal spin pairing,
Phys. Rev. Lett. {\bf 103}, 057003  (2009).

\bibitem{VollhardtWolfle1990}
D. Vollhardt  and P.  W\"olfle,
{\it The superfluid phases of helium 3} (Taylor and Francis, London, 1990).

 \bibitem{Volovik1999}
G.E. Volovik,  
Fermion zero modes on vortices in  chiral superconductors,  
JETP. Lett. {\bf 70}, 609--614 (1999); cond-mat/9909426.

\bibitem{ReadGreen2000}
N. Read and D. Green,
Paired states of fermions in two dimensions with breaking of parity and time-reversal symmetries and the fractional quantum Hall effect,
Phys. Rev. B {\bf 61}, 10267--10297 (2000).

\bibitem{Ivanov2001}
D.A. Ivanov,
Non-abelian statistics of half-quantum vortices in p-wave superconductors,
Phys. Rev. Lett. {\bf 86}, 268 (2001).

\bibitem{Thuneberg1986}
E. V. Thuneberg,
Identification of vortices in superfluid $^3$He-B,
Phys. Rev. Lett. {\bf 56}, 359--362 (1986).

\bibitem{VolovikSalomaa1985}
G.E. Volovik and M.M. Salomaa, 
Spontaneous breaking of axial symmetry in $v$-vortices in superfluid $^3$He-B,
JETP. Lett. {\bf 42}, 521--524  (1985).

\bibitem{Kondo1991}
Y. Kondo,  J.S. Korhonen, M. Krusius, V.V. Dmitriev,  Yu. M.  Mukharskiy, E.B. Sonin and G.E. Volovik, 
Direct observation of the nonaxisymmetric vortex in superfluid $^3$He-B,
Phys. Rev. Lett.  {\bf 67}, 81--84 (1991).

\bibitem{Volovik1990}
G. E. Volovik,
Half quantum vortices in the B phase of superfluid $^3$He,
JETP. Lett. {\bf 52}, 358 (1990).

\bibitem{Silaev2015}
M.A. Silaev, E.V. Thuneberg, and M. Fogelstr\"om,
Lifshitz Transition in the Double-Core Vortex in $^3$He-B,
Phys. Rev. Lett. {\bf  115}, 235301 (2015).

\bibitem{MineyevVolovik1978}
V.P. Mineyev, G.E. Volovik, 
 Planar and linear solitons in superfluid $^3$He,
Phys. Rev. B {\bf 18}, 3197--3203 (1978).

\bibitem{Mermin1979} 
N.D. Mermin,
The topological theory of defects in ordered media,
Rev. Mod. Phys. {\bf 51}, 591 (1979).

\bibitem{Michel1980} 
L. Michel,
Symmetry defects and broken symmetry. Configurations Hidden Symmetry
Rev. Mod. Phys. {\bf 52}, 617 (1980).

\bibitem{Kondo1992} 
Y. Kondo,  J.S. Korhonen, M. Krusius, V.V. Dmitriev, E.V. Thuneberg and  G.E. Volovik, 
 Combined spin - mass vortices with soliton tail in superfluid $^3$He-B, 
Phys. Rev. Lett. {\bf 68}, 3331 (1992).
 
\bibitem{Seji2019}
Seji Kang, Sang Won Seo, Hiromitsu Takeuchi, and Y. Shin,
Observation of Wall-Vortex Composite Defects in a Spinor Bose-Einstein Condensate,
Phys. Rev. Lett. {\bf 122}, 095301 (2019).

\bibitem{Liu2020}
I-Kang Liu, Shih-Chuan Gou, H. Takeuchi,
Phase diagram of solitons in the polar phase of a Spin-1 Bose-Einstein condensate,
arXiv: 2002.06088.

\bibitem{Volovik1978}
G.E. Volovik, 
Topological singularities on the surface of an ordered system,
Pis'ma. Zh. Eksp. Teor. Fiz. {\bf 28}, 65--68 (1978); JETP. Lett. {\bf 28} 59--62 (1978).

\bibitem{Mermin1981}
N.D. Mermin,
E Pluribus Boojum: the physicist as neologist,
Physics Today {\bf 34}, 46 (1981).

\bibitem{MMV1978}
N.D. Mermin, V.P. Mineev, G.E. Volovik, 
Topological analysis of cores of singularities in the $^3$He-A,
J. Low Temp. Phys.  {\bf 33}, 117--126 (1978).

\bibitem{Cornwall1999}
J.M. Cornwall,
Center vortices, nexuses, and the Georgi-Glashow model,
Phys. Rev. D {\bf 59}, 125015 (1999).

\bibitem{GoloMonastyrky1978}
V.L. Golo and M.I. Monastyrsky,
Gauge groups and topological invariants of vacuum manifolds,
Ann. l’I.H.P. Phys. theorique, {\bf 28}, 75–89 (1978).

\bibitem{Volovik1996}
G.E. Volovik, 
Glass state of superfluid $^3$He-A in aerogel, 
Pis'ma. Zh. Eksp. Teor. Fiz. {\bf  63},  281--284 (1996);
JETP Letters {\bf 63},  301--304 (1996 );  
cond-mat/9602019.

\bibitem{Dmitriev2010}
V.V. Dmitriev, D.A. Krasnikhin, N. Mulders, A.A. Senin, G.E. Volovik and A.N. Yudin,
Orbital glass and spin glass states of  $^3$He-A  in aerogel,
Pis'ma. Zh. Eksp. Teor. Fiz. {\bf 91}, 669--675 (2010);  
JETP Lett. {\bf 91}, 599--606 (2010);
arXiv:1004.5483.

\bibitem{Glasses2019}
G.E. Volovik, J. Rysti, J.T. M\"akinen, V.B. Eltsov,
Spin, orbital, Weyl and other glasses in topological superfluids,
J. Low Temp. Phys. {\bf 196}, 82--101 (2019).


\bibitem{VolovikMineev1977} 
G.E. Volovik, V.P. Mineev, 
Investigation of  singularities in superfluid $^3$He and liquid crystals by homotopic topology methods,
JETP {\bf 45} 1186--1196 (1977).

\bibitem{Schwarz1982}  
A.S. Schwarz, 
Field theories with no local conservation of the electric charge,
Nucl. Phys. B {\bf 208},  141--158 (1982).

\bibitem{SalomaaVolovik1988}
 M.M. Salomaa,  G~E. Volovik,
Cosmiclike domain walls in superfluid $^3$He-B: Instantons and diabolical points in $({\bf k},{\bf r})$ space, 
Phys. Rev. B {\bf 37}, 9298--9311 (1988).


\bibitem{Nash1988} 
Charles Nash, Siddhartha Sen, 
{\it Topology and Geometry for Physicists} (Academic Press, 1988).

\bibitem{Surovtsev2007}
E.V. Surovtsev and I.A. Fomin, 
Model Calculation of Orientational Effect of Deformed Aerogel on the Order Parameter of Superfluid $^3$He, 
J. Low. Temp. Phys. {\bf 150}, 487  (2008).

\bibitem{Rainer1977}
D. Rainer, M. Vuorio, 
Small objects in superfluid $^3$He, 
J. Phys. C {\bf 10}, 3093 (1977). 


\bibitem{volovik2000}
G. E. Volovik, 
Monopoles and fractional vortices in chiral superconductors, 
Proc. Natl. Acad. Sci. U.S.A. {\bf 97}, 2431--2436 (2000).


\bibitem{Blaha1976}
S. Blaha,
Quantization rules for point singularities in superfluid 3He and liquid crystals,
Phys. Rev. Lett. {\bf 36}, 874--876 (1976).

\bibitem{Volovik1976}
G.E. Volovik, V.P. Mineev,
Vortices with free ends in superfluid 3He-A,
JETP. Lett. {\bf 23}, 593--596 (1976).

\bibitem{Saurabh2019}
A. Saurabh and T. Vachaspati,
Monopole–antimonopole: interaction, scattering and creation,
Phil. Trans. R. Soc. A {\bf 377}, 20190143 (2019).

\bibitem{Shafi2019}
G. Lazarides and Q. Shafi,
Monopoles, Strings, and Necklaces in SO(10) and $E_6$,
J. High Energ. Phys. {\bf 2019}, 193 (2019).

\bibitem{Volovik2019d}
G.E. Volovik,
Composite topological objects in topological superfluids,
arXiv:1912.05962.

\bibitem{skyrme1962}
T.H.R. Skyrme, 
A unified field theory of mesons and baryons, 
Nucl. Phys {\bf 31}, 556--569 (1962).

\bibitem{khawaja2001}
U.A. Khawaja, H. Stoof, 
Skyrmions in a ferromagnetic Bose-Einstein Condensate, 
Nature {\bf 411}, 918-920 (2001).

\bibitem{MerminHo1976}
N. Mermin and T.-L. Ho, 
Circulation and angular momentum in the 
A-phase of superfluid helium-3,
Phys. Rev. Lett. {\bf 36}, 594 (1976).

\bibitem{Anderson1977}
P. W. Anderson and G. Toulouse,
Phase slippage without vortex cores: vortex textures in superfluid $^3$He,
Phys. Rev. Lett. {\bf 38}, 508--511 (1977).

\bibitem{Chechetkin1976}
V.R. Chechetkin,
Types of vortex solutions in superfluid $^3$He,
JETP {\bf 44}, 766--772 (1976).

\bibitem{Volovik1977}
G.E. Volovik, N.B. Kopnin,
On the rotating $^3$He- A,
Pis'ma. Zh. Eksp. Teor. Fiz. {\bf 25}, 26--28 (1977); JETP Lett. {\bf 25}, 22--24 (1977).

\bibitem{Seppala1984}
H.K. Seppälä, P.J. Hakonen, M. Krusius, T. Ohmi, M.M. Salomaa, J.T. Simola, and G.E. Volovik,
Continuous vortices with broken symmetry in rotating superfluid $^3$He-A,
Phys. Rev. Lett. {\bf 52}, 1802--1805 (1984).

\bibitem{Pekola1990}
J.P. Pekola, K. Torizuka, A.J. Manninen, J.M. Kyynäräinen and G.E. Volovik,
Observation of a topological transition in the $^3$He-A vortices,
Phys. Rev. Lett {\bf 65}, 3293–3296 (1990).

\bibitem{Bevan1997}
T.D.C. Bevan, A.J. Manninen, J.B. Cook, J.R. Hook, H.E. Hall, T. Vachaspati and G.E. Volovik,
Momentum creation by vortices in superfluid 3He as a model of primordial baryogenesis,
Nature {\bf 386}, 689-692 (1997).

\bibitem{Samme2019}
S.M. Dahir, A.F. Volkov, and I.M. Eremin,
Interaction of skyrmions and Pearl vortices in superconductor-chiral ferromagnet heterostructures,
Phys. Rev. Lett. {\bf 122}, 097001 (2019).

\bibitem{Stefan2019}
S. Rex, I.V. Gornyi, and A.D. Mirlin,
Majorana bound states in magnetic skyrmions imposed onto a superconductor,
Phys. Rev. B {\bf 100}, 064504 (2019).

\bibitem{Vorontsov2007}
A.B. Vorontsov and J. A. Sauls,
Crystalline order in superfluid $^3$He films,
Phys. Rev. Lett. {\bf 98}, 045301 (2007).

\bibitem{Levitin2019}
L.V. Levitin, B. Yager, L. Sumner, B. Cowan, A.J. Casey, J. Saunders, N. Zhelev, R.G. Bennett, and J.M. Parpia,
Evidence for a spatially modulated superfluid phase of $^3$He under confinement,
Phys. Rev. Lett. {\bf 122}, 085301 (2019).

\bibitem{Shook2019}
 J. Shook, V. Vadakumbatt, P. Senarath Yapa, C. Doolin, R. Boyack, P.H. Kim, G.G. Popowich, F. Souris, H. Christani, J. Maciejko, J.P. Davis,
Stabilized pair density wave via nanoscale confinement of superfluid $^3$He-A,
Phys. Rev. Lett.  {\bf 124}, 015301 (2020).
\end{thebibliography}
\end{document}